\documentclass[12pt]{article}
\usepackage{epsfig}

\makeatletter
\@addtoreset{equation}{section}

\newcommand{\p}{\partial}
\newcommand{\K}{{\cal K}}
\newcommand{\T}{{T_{\rm cl}}}

\begin{document}

\begin{titlepage}
\begin{center}

\hfill\parbox{4cm}{
{\normalsize\tt hep-th/0209232}\\
{\normalsize TIT/HEP-484}\\
{\normalsize UT-Komaba/02-08}
}

\vskip 1in

{\LARGE
\bf Brane - Antibrane as a Defect of \\[3mm]
Tachyon Condensation}

\vskip 0.3in

{\large
Koji {\sc Hashimoto} $^a$\footnote{\tt koji@hep1.c.u-tokyo.ac.jp} 
and Norisuke {\sc Sakai} $^b$\footnote{{\tt nsakai@th.phys.titech.ac.jp}}
}

\vskip 0.15in

${}^a$ {\it Institute of Physics, University of Tokyo, Komaba}\\
{\it Tokyo 153-8902, Japan}\\[3pt]
${}^b$ {\it Department of Physics, Tokyo Institute of Technology}\\
{\it Tokyo 153-8551, Japan}\\[0.3in]

{\normalsize September, 2002}

\end{center}

\vskip .3in


\begin{abstract}
\normalsize\noindent 
 In a tachyon effective field theory of a non-BPS brane, we construct a 
 classical solution representing a parallel brane-antibrane. The 
 solution is  made of a kink and an antikink placed at antipodal points 
 of $S^1$. 
 Estimation of the brane energy suggests an excitation of
 a string connecting the two branes,  even though the theory is
 Abelian. We performed fluctuation analysis around the obtained
 solution, and find the structure of the supersymmetry breaking by the
 co-existence of the brane and the antibrane. We discuss possible
 processes of the pair-annihilation of the brane defect.
\end{abstract}

\vfill

\end{titlepage}
\setcounter{footnote}{0}

\pagebreak
\renewcommand{\thepage}{\arabic{page}}
\pagebreak


\section{Introduction and setup}

After the success of K-theory in string theory, 
a standpoint that various D-brane configurations can be obtained by the
open string tachyon condensation is established. This viewpoint is the
third one to provide the description of D-branes, as well as the
conformal field theory approach and the supergravity description. 
Around this new possibility, there are two currents of study, one is 
the verification of Sen's conjecture \cite{Sen,sen2,sen3} 
on tachyon condensation with use of
string field theories, and another is to utilize this new description to
describe off-shell dynamics of D-branes including time-dependent
backgrounds and application to realistic braneworld models.

Our attempt in this note is oriented rather to the second point above.
The aim of this note is to see how far we can proceed for constructing
brane configurations as a defect of the tachyon condensation in string
theory,\footnote{The early studies of open string tachyon dynamics, see
\cite{Halpern}.} by restricting our attention to the brane-antibrane
configurations. It is known that,  to obtain a desired brane
configuration as a tachyon condensate,  one has to prepare an
appropriate number of non-BPS branes from the first place.  For example,
if one wants to construct a solution of two parallel D-branes, one has
to prepare at least two unstable branes of higher dimensional
worldvolume. Therefore the inter-brane interactions of the resultant
defect branes are already encoded in the original non-Abelian setup.  
Then one can extract from the solution 
the off-diagonal interaction which corresponds to
the excitation of the string connecting two branes, and it was found in
\cite{Minahan} that this excitation reproduces the lowest mass squared
of the string excitation.\footnote{Another interesting example is
  non-commutative solitons in an Abelian worldvolume gauge theory of
  D-branes \cite{aga}. In this case because of the non-commutativity the
  worldvolume theory effectively becomes non-Abelian $U(\infty)$, and
  off-diagonal modes in the sectors of the solitons
  (lower-dimensional D-branes bounded) reproduce the spectra of
  strings stretched between them.  }

However, we will show explicitly 
that a solution representing brane and antibrane is
possible in an Abelian setup. 
Since this brane-antibrane should be a kink-antikink (or
rather to say, wall-antiwall \cite{Maru:2000sx}), 
this can be constructed by a
simple Abelian model with only one tachyon field. 
The intriguing aspect here is that 
non-Abelian structure is expected to emerge from the Abelian
model. 

Another motivation for the construction of the brane-antibrane solution
in superstring tachyon models is to realize in string theory 
the outcome of the field theoretical analysis in non-linear sigma
models,\footnote{An interesting example was provided in \cite{hirano}, 
where
the kinky-lump solution in a non-linear sigma model constructed in 
\cite{Gauntlett:2000de} was realized as a D-brane configuration 
made of superstring tachyon condensation. 
} phenomenology with the extra dimensions, and braneworlds. 
In particular, brane-antibrane system breaks supersymmetries, which is
phenomenologically preferable. We perform the fluctuation analysis of
the constructed solution, and investigate how the spectra with
supersymmetry breaking appears in the superstring tachyon model. 
In accordance to the recent development on the tachyon cosmology, we
also study the time-dependent decay process of the brane and antibrane.

For concreteness we shall use the two-derivative truncation of the
boundary string field theory (BSFT) action \cite{Witten1,moore} 
for the field theory of
the tachyon on  the non-BPS D-brane. Section 2 provides a
brane-antibrane solution of this theory, and in section 3 the force
between the branes is estimated to suggest the
off-diagonal string interaction. In section 4 we study the fluctuation
spectra and find supersymmetry breaking structure. Section 5 gives the
decay width of the brane and antibrane. Conclusion and discussions are
provided in the final section. In Appendix A we find similar solutions
exist also in the BSFT with higher derivative corrections.

\section{Brane-antibrane solution}

\subsection{BSFT action and its solutions}

As mentioned in the introduction, we 
use the two-derivative truncation of the
BSFT action of a non-BPS brane \cite{moore}, 
some of which are called Minahan-Zwiebach (MZ) model \cite{MZ,MZg}.
This kind of BSFT model exhibits very interesting properties which
are consistent with string theory. In particular, 
exact classical solutions representing a D-brane exist, 
and they have appropriate Ramond-Ramond (RR) charges. Furthermore, 
the fluctuation spectra around the soliton solutions 
are exactly solvable, and found to be consistent with 
D-brane spectra \cite{MZ, MZg, Hashimoto:2002xe}. 

The MZ model action for a single non-BPS D$(p+1)$-brane is given by 
\begin{eqnarray}
S = -\int\! d^{p+2}x\; (\K')^2 \left(1 + \p_\mu T \p^\mu T\right),
\label{Lag}
\end{eqnarray}
where $\K'(T)$ is a function of $T$ which specifies the tachyon
potential. For the action obtained by the derivative truncation of the
BSFT, we have
\begin{eqnarray}
 \K' = \exp(-T^2/2).
\label{bsftpot}
\end{eqnarray}
In the action (\ref{Lag}), 
for simplicity we ignored overall factor of the non-BPS brane 
tension. 
In Sec.\ 2 and 3,  we neglect the dependence on the coordinate 
other than one world volume direction $x$. 
The dimension of the  tachyon field is
appropriately normalized with $\alpha'$. 

A characteristic point of this action is that the potential term 
$(\K')^2$ is multiplied also on the kinetic term. 
For the BSFT potential $(\K')^2 = {\rm e}^{-T^2}$, a perturvative vacuum $T=0$
is meta-stable, and there are two  
true vacua, $T=\pm \infty$.
The equations of motion are given by
\begin{eqnarray}
 \partial_\mu \partial^\mu T - \frac{\K''}{\K'} \left(1-
 \partial_\mu T \partial^\mu T\right)=0, 
\label{eqm}
\end{eqnarray}
provided that $\K' \neq 0$.
It is known that the above Lagrangian allows a very simple linear
solution 
\begin{eqnarray}
T=x-x_0 
\label{linear}
\end{eqnarray}
which represents a D$p$-brane, since 
according to Sen's conjecture the kink solution which interpolates
two vacua ($T=\pm \infty$) is a BPS D-brane with codimension-one world
volume.  

The location of the D-brane is determined by the equation $T(x)=0$,
since the point $T=0$ is just the point between the two true vacua,
and any kink solution should pass this mid point. Therefore, the above
single kink solution represents a BPS D$p$-brane located at $x=x_0$.
It should be noted that another solution $T=-(x-x_0)$ denotes an  
anti-${\mbox{D$p$}}$-brane, since the coupling to the RR gauge field 
$\int {\rm e}^{-T^2}dT \wedge C $ gives negative contribution. Therefore, if
the solution passes  zero of $T$ from negative to positive it represents
a BPS D-brane while if it passes in the opposite way it is an
antibrane.  

\subsection{Brane-antibrane solutions}

Let us proceed to obtain new solutions representing a pair of a brane and
an antibrane.  
What we require for 
a solution of the brane and antibrane 
is that it satisfies the following properties: 
\begin{itemize}
 \item At both of the spatial infinities $x=\pm \infty$, 
T should be located at
       near one of the true vacuum : $T=\infty$.
\item The equation $T(x)=0$ with the solution $T$ should have two
       solutions for $x$, which are the locations of the brane and 
       the antibrane. $\partial_x T$ has different signs at these zeros of
       the solution $T$.
\end{itemize}
Let us find a solution satisfying these properties. 
In fact, it is possible to construct a general solution 
for the equations of motion (\ref{eqm}) if we restrict our attention to
a field configuration which depends only one direction $x$. 
By multiplying $\p_x T$ on (\ref{eqm}), 
the equation of motion is recast into the form 
\begin{eqnarray}
 \frac{\K''}{\K'} \p_x T = \frac{\p_x T}{1-(\p_x T)^2} \p_x^2 T.
\end{eqnarray}
This can be integrated by $x$, and the result is 
\begin{eqnarray}
 -\log \K' = \frac12 \log \left( 1-(\p_x T)^2\right) + \mbox{const.}
\end{eqnarray}
Thus we obtain a relation
\begin{eqnarray}
 (\p_x T)^2 = 1-A (\K')^{-2},
\label{relation}
\end{eqnarray}
where $A$ is an integration constant parameter.
One observes that the vanishing $A$ gives the 
simple linear solution (\ref{linear}). 
Therefore this parameter $A$ is the one which characterizes the
deviation from the BPS single kink solution toward the 
brane-antibrane solution.  
The physical parameter of the brane and antibrane which is
expected is the inter-brane separation, 
and we will relate this to the above $A$. 
Before going there, the following boundary condition will be found
to be useful: 
\begin{eqnarray}
 T=T_0, \quad \p_x T=0 \quad \mbox{at} \quad x=x_0.
\end{eqnarray}
Then, for the case of BSFT potential (\ref{bsftpot}), 
 we have $A = {\rm e}^{-T_0^2}$, and integration of the equation
(\ref{relation}) gives an explicit expression for the general solution: 
\begin{eqnarray}
 x = x_0 \pm 
{\rm e}^{T_0^2/2}\int_{T_0}^{T} 
\frac{1}{\sqrt{{\rm e}^{T_0^2}-{\rm e}^{T^2}}} dT. 
\label{solution1}
\end{eqnarray}
The BPS limit is $T_0 \rightarrow \infty$, though the above expression
becomes meaningless in that limit. We will find later that this is
because the brane and the antibrane are pushed out to the spatial
infinities in this limit.

\subsection{Properties of the solution}

Here we shall show that the solution (\ref{solution1}) is actually a
solution representing a brane-antibrane pair. 
In the following of this section and the next section, we consider only
the BSFT potential $(\ref{bsftpot})$ for simplicity. 
Let us suppose that $T_0<0$ and $x_0=0$. This does not lose any
generality since the original system is invariant under 
$T \rightarrow -T$ and also translationally invariant. 

Interestingly, the solution (\ref{solution1}) can be consistent only if 
$|T| <|T_0|$. So let us study what happens around the critical point
$T\sim \pm T_0$. For the tachyon field near the critical point
the integrand  in the
solution (\ref{solution1}) seems to be dangerous because of a possible 
divergence. 
In order to check the behavior around the critical point, 
we expand the tachyon as 
\begin{eqnarray}
 T = T_0+\delta 
\end{eqnarray}
where positive $\delta$ is taken to be infinitesimally small. 
Then the integral is estimated as
\begin{eqnarray}
x_1-x = \int_{T_0-\delta}^{T_0} \frac{dT}{\sqrt{1-{\rm e}^{T^2-T_0^2}}}
\sim
 \int_0^\delta \frac{1}{\sqrt{2|T_0|\delta'}} d \delta' 
=\sqrt{\frac{2\delta}{|T_0|}}. 
\end{eqnarray}
Thus the integral is finite and has no divergence.  
This means that if we plot the solution in the $x$-$T$ space 
the solution is a curved line with a finite length, 
connecting $(x=0, T=T_0)$
and $(x=x_1, T=-T_0)$. In fact, the second term in (\ref{solution1})
is convergent for $T=-T_0$, thus there is the end of the base space at
$x=\pm x_1$ where
\begin{eqnarray}
 x_1 \equiv 
{\rm e}^{T_0^2/2}\int_{T_0}^{-T_0} 
\frac{1}{\sqrt{{\rm e}^{T_0^2}-{\rm e}^{T^2}}} dT. 
\end{eqnarray}
How should this finiteness be understood ? 
One idea to interpret this strange property 
is to assume that the solution lives in a compact space, 
$-x_1 \leq x < x_1$. If one assumes this, the size of the
compactified dimension ($2x_1$) determines 
the parameter $T_0$ through the above equation. 
The other idea is that one can have a periodic solution,
ends up with an infinite repetition of the solution. Here the
period is of course the same as the first case, $2x_1$.

It should be noted that the solution $T(x)$
has two zeros in the definition
region $-x_1\leq x<x_1$. 
This is almost obvious since substituting $T=0$ in
(\ref{solution1}) gives $x=\pm x_1 /2$, because the integrand is
symmetric under $T \rightarrow -T$. 
The zeros of the tachyon profile represent the location of the
D-branes.
The tachyon potential 
${\rm e}^{-T^2}$ has two vacua, $T=\pm \infty$, so there are two types of
topological defects: 
a kink and an anti-kink, with regard to its orientation. 
In our case, $\p_x T (x_1/2)$ is positive while 
$\p_x T (-x_1/2)$ is negative, hence our solution (\ref{solution1})
is actually a pair of a kink and an antikink. 
Therefore, we conclude that this
solution represents the brane-antibrane pair. 
The inter-brane separation is $x_1$.

\begin{figure}[tdp]
\begin{center}
\begin{minipage}{100mm}
\begin{center}
   \leavevmode
   \epsfxsize=70mm
   \epsfbox{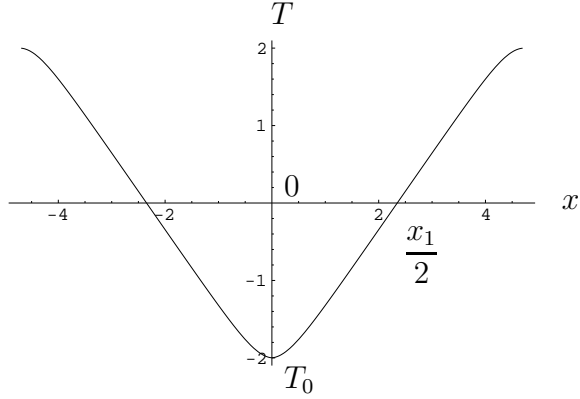}
\put(10,60){$x$}
\put(-50,40){$\displaystyle\frac{x_1}{2}$}
\put(-100,130){$T$}
\put(-95,65){$0$}
\put(-95,-8){$T_0$}
\caption{Solution (\ref{solution1}) with $x_0=0$ and $T_0 = -2$.
}
\label{fig1}
\end{center}
\end{minipage}
\end{center}
\end{figure}

We plot a numerical  solution in Fig.\ \ref{fig1}. One can easily see
that most part of the solution is linear, and the edge of the lines
(at $x=0,\pm x_1$) are smoothly connected. The kink and the antikink are 
joined. 

For the later purpose, we estimate the brane separation in terms of
$T_0$. Suppose that we have very large $|T_0|$. Then the integrand in the
solution (\ref{solution1}) is approximated by a constant 
in almost all the region of $T$. This constant is actually given by the
value at $T=0$, as 
\begin{eqnarray}
 \frac{\p T}{\p x}\biggm|_{T=0} = \pm q
\label{value}
\end{eqnarray}
where $q\equiv \sqrt{1-{\rm e}^{-T_0^2}}$.
Therefore, we can estimate that in this large $|T_0|$ limit
the brane separation $x_1$ can be given by 
\begin{eqnarray}
 x_1 \equiv 
{\rm e}^{T_0^2/2}\int_{T_0}^{-T_0} 
\frac{1}{\sqrt{{\rm e}^{T_0^2}-{\rm e}^{T^2}}} dT 
\sim 
\int_{T_0}^{-T_0} 
\frac{1}{q}\;
dT
=
\frac{2|T_0|}{q}.
\end{eqnarray}
Therefore, approximately the brane separation $x_1$ is given by $2|T_0|
+{\cal O} ({\rm e}^{-T_0^2})$. 

Summing up the above estimations, an approximate form of the solution 
for positive $x$ is given as
\begin{eqnarray}
T \sim \left\{
 \begin{array}{ll}
-|T_0|+\frac{|T_0|}{2}x^2
&
\hspace{10mm}
0 < x < \frac{q}{|T_0|} \\
q \left(x-x_1/2\right) 
& 
\hspace{10mm}
\frac{q}{|T_0|}<x<\left(x_1-\frac{q}{|T_0|}\right)
\\
|T_0|-\frac{|T_0|}{2}(x-x_1)^2 
& 
\hspace{10mm}
\left(x_1-\frac{q}{|T_0|}\right)<x<x_1
 \end{array}
\right.
\label{approx}
\end{eqnarray}

\section{Force between the branes}

It is obvious that since this solution is a wall antiwall pair, 
the energy is reduced if the distance between the two approaches zero
while the compactification radius fixed. However, these two parameters
are related for the obtained solution, of course. 
The obtained configuration is meta-stable because the brane and the
antibrane are located in an antipodal manner. 
The instability   
associated with the off-shell deformation of decreasing the inter-brane
distance is the tachyonic instability which comes 
from the interaction between the two.

In string theory, naively this interaction mode is a closed string mode
such as gravity and RR-interaction which are attractive force between
the brane and the antibrane. However,
from the first setup we neglected the closed string channel, so the
appearance of the gravity interaction is hardly expected
here.

Let us recall that the gravity and the RR interaction 
are arranged into the
one loop amplitude of open string stretched between the two, if one sums
up all the massive modes of closed strings. This is the 
s-t channel duality. 
Now in the dual channel (open string) description, 
we prepared only the lowest modes (tachyons), that is why we cannot
expect the reproduction of the gravity interaction. 
Then, what is the interaction of branes in our case? 
Instead of the closed string interaction, 
we may expect only the lowest interaction of stretched open
string. This is a massive state of the mass $\sim x_1$.

The above argument shows that the inter-brane interaction is now
dictated by the massive state with the mass $\sim x_1$, not by the
gravity interaction. This should be seen in the solution constructed in
this paper. For that purpose, let us evaluate the potential between the
branes. 
Denoting 
the energy of the original single brane $T=x$ as 
 $E_0$, 
\begin{eqnarray}
 E_0 \equiv \int_{-\infty}^\infty\! dx \; 
\left[{\rm e}^{-T^2} \left(
1+({\p_x T})^2
\right)
\right]_{T=x} 
=2\int_{-\infty}^\infty\! dx\; {\rm e}^{-x^2}, 
\end{eqnarray}
we can evaluate the potential energy by an excess energy $\Delta E$ 
of the brane-antibrane configuration compared to twice of the energy $E_0$ 
of the single isolated brane, 
\begin{eqnarray}
 \Delta E = 
\int_{-x_1}^{x_1}\!dx\; {\rm e}^{-T^2} \left(
1+({\p_x T})^2
\right) 
-2 E_0 .
\label{energy}
\end{eqnarray}
The multiplication factor 2 in the last term 
is necessary since we have two branes. 

By using (\ref{solution1}) with $x_0=0$, we change integration variable 
from $x$ to $T$ and obtain 
\begin{eqnarray}
 \Delta E 
 &\!\!
 = 
 &\!\!
2\int_{T_0}^{-T_0}\!dT\; \frac{2{\rm e}^{-T^2}-{\rm e}^{-T_0^2}}{ 
\sqrt{1-{\rm e}^{T^2-T_0^2}}} 
-4 \int_{-\infty}^{\infty}\!dT\; {\rm e}^{-T^2}
 \\
 &\!\!
 = 
 &\!\!
8\left[\int_{0}^{|T_0|}\!dT\; {\rm e}^{-T^2}\left(\frac{1}{ 
\sqrt{1-{\rm e}^{T^2-T_0^2}}} -1\right)
-\int_{|T_0|}^\infty dT {\rm e}^{-T^2}
- \int_{0}^{|T_0|}\!dT\; 
\frac{{\rm e}^{-T_0^2} }{ 2\sqrt{1-{\rm e}^{T^2-T_0^2}}}
\right]
\nonumber
\label{energy2}
\end{eqnarray}
We can see that each term in the last line gives contributions of order 
${\cal O}({\rm e}^{-T_0^2})$ with possible powers of $T_0$. 
Since the brane-antibrane distance is $x_1 \sim 2|T_0|$, we obtain 
\begin{eqnarray}
 \Delta E \sim -\frac{1}{x_1^n} {\rm e}^{-x_1^2/4} + \mbox{higher} \; \; 
 \mbox{orders}.
\end{eqnarray}
Here we have not determined the precise value of the integer $n$ 
(this can be negative), since the
important part is the exponential. 
If we regard this as a Yukawa-type
potential 
\begin{eqnarray}
 V(x_1) = \frac{e_1 e_2}{x_1^n} {\rm e}^{-M x_1}
\label{yukawa}
\end{eqnarray}
where $M$ is the typical mass of the particle which causes the
interaction, we see 
\begin{eqnarray}
 M \sim x_1, 
\end{eqnarray}
up to a normalization.\footnote{ 
The above argument does not completely ensure that the mass is actually
linear in $x_1$, since the mass itself is expected to depend on the
separation $x_1$ and in principle one cannot extract the mass dependence
only from the excess energy.} 
Here we see the appearance of a new mass scale $\sim x_1$. Since the
interbrane interaction should come from strings connecting the branes,
this suggests that the typical mass scale for these strings should be
of order $\sim x_1$. This is consistent with the string theory mass
spectra. However, the string connecting the two propagates along the
branes while the particle producing the potential (\ref{yukawa})
propagates transversely. Although this difference might be explained 
by the duality, precise reproduction of the
corresponding string excitation is yet to be found.

\section{Fluctuation analysis}

In this section we investigate the low energy effective field theories 
induced on the brane and antibrane. We do not consider the effective 
theories on each brane separately,  since there exists weak 
interactions between the branes, as we have seen in the previous
section. Because of this weak interaction, the degenerate mass spectra
on each brane resolves to form a fine structure. This resolution
is actually due to the supersymmetry breaking, since a single brane 
(a single wall) itself is supersymmetric and is a BPS state while the
coexistence  of the brane and the antibrane breaks the supersymmetry. 
Since the inter-brane interaction as well as 
the supersymmetry breaking are 
exponentially suppressed as a function of the distance between 
the brane and antibrane, 
 the resulting tower of boson and fermion 
spectra exhibits fine splittings. 

\subsection{Tachyon fluctuation}

First let us derive the Lagrangian for the fluctuation of the tachyon
fields around the classical solution $\T$ (\ref{solution1}). We expand
the field as
\begin{eqnarray}
T(x, y^{\hat{\mu}}) =  \T (x) + t(x,y^{\hat{\mu}}),
\end{eqnarray}
where $y$ denotes the defect brane worldvolume coordinate,
$\hat{\mu}=0,1,\cdots,p-1$. Substituting this expression to the original
Lagrangian (\ref{Lag})
and pick up the terms quadratic in $t$, we obtain 
\begin{eqnarray}
 {\cal L}_t = 
-\left(\K'\right)^2
\left[
\left(\left(\frac{\K''}{\K'}\right)^2 + \frac{\K'''}{\K'}\right)
(1 + (\p_x \T)^2)  t^2  + \frac{\K''}{\K'}4  \p_x \T t \p_x t + 
(\p_x t)^2 + (\p_{\hat{\mu}}t)^2
\right].
\label{flua}
\end{eqnarray}
Here and in the following, $\K$ is the function into which the classical
solution $\T$ is substituted.
Following the technique developed in \cite{MZ,MZg}, we make a field
redefinition as 
\begin{eqnarray}
 \hat{t} \equiv  \K'(\T) \; t.
\label{redef}
\end{eqnarray}
Then after a partial integration we obtain
\begin{eqnarray}
 {\cal L}_t = -\hat{t} 
\left[
\left(-\p_x^2 + V_{\rm tachyon}(\T)\right)
-\p_{\hat{\mu}}^2
\right]\hat{t}, 
\qquad 
V_{\rm tachyon}(\T)\equiv \frac{\K'''}{\K'}.
\end{eqnarray}
Therefore, if we expand the fluctuation fields as
\begin{eqnarray}
 t = \sum_n t_n(y^{\hat{\mu}}) u_n(x)
\end{eqnarray}
in which the function $u_n(x)$ solves the eigenvalue equations
\begin{eqnarray}
 \left(
-\p_x^2 + V_{\rm tachyon}(\T)
\right) u_n (x)= m_n^2 u_n(x),
\label{flucL}
\end{eqnarray}
the corresponding mode $\hat{t}_n$ has the mass $m_n$, 
\begin{eqnarray}
\p_{\hat{\mu}}^2 \hat{t}_n(y^{\hat{\mu}}) = 
m_n^2 \hat{t}_n(y^{\hat{\mu}}) . 
\end{eqnarray}

For the BSFT potential (\ref{bsftpot}), 
$ V_{\rm tachyon}(\T)$ is 
given by $\T^2-1$. Therefore if one has a single D-brane solution as a
background, the potential
becomes harmonic and the Schr\"odinger equation (\ref{flucL}) becomes 
solvable, results in a spectrum with equal
spacing.

Since we have two branes, there should be two almost massless modes
as the low energy modes. Among them, we may expect that 
there is a massless 
Nambu-Goldstone (NG) boson coming from the breaking of the translational
symmetry of the original system.  
The other modes should be tachyonic, since this scalar mode should be 
related to the inter-brane interaction which was shown to be a negative
mode in the previous section. Let us see this in more detail using the
BSFT potential (\ref{bsftpot}).

In fact, the potential which fluctuation modes feel is the double-well
type. The potential $V(\T)$ has two
minima at the zeros of the solution $\T$, $x=\pm x_1/2$. 
If the inter-brane separation is large enough, 
we may expect massless NG modes for each brane. Now, if the separation
is not so large, these two NG modes are not really a NG mode, and they
interact with each other. A basic knowledge on the quantum mechanics
shows that the symmetric wave function is the true lowest mode while the
antisymmetric wave function is the first excited level. Thus we expect
that this symmetric one is tachyonic and the antisymmetric one is 
the massless NG mode. 

The correct expression for the massless NG mode is well-known, 
\begin{eqnarray}
 t(x) = \p_x \T.
\label{ng}
\end{eqnarray}
It is easy to check that this satisfies (\ref{flucL}) with $m^2 =0$, if
we use the equations of motion for $\T$, (\ref{eqm}). And one can
check that this NG mode is actually antisymmetric under the parity
transformation $x \leftrightarrow -x$. The expression (\ref{solution1}) 
of the solution shows $\T$ is an even function of $x$, thus its
derivative $\p_x \T$ is an odd function (antisymmetric with respect
to the double well potential). Fig.\ 2 is the plot of the NG mode. 

One observes that the NG mode is almost flat, as is obvious from the
expression (\ref{ng}) and the fact that the solution (\ref{solution1})
is almost linear around the bottom of the potential. 
In fact, the single kink solution is a simple linear function $\T=x$,
thus its NG mode is just constant. To describe in a different manner, 
we see that the potential for the fluctuation (\ref{flucL}) is that of
a harmonic oscillator for $\T=x$, thus the lowest mode is Gaussian,
$\hat{t} = A \exp(-x^2)$. This shows that through the field redefinition
(\ref{redef}) the fluctuation $t$ is constant. 

The massless NG mode is an antisymmetric combination of the flat
configuration, thus we expect that the tachyonic one is a symmetric
combination, that is, almost constant everywhere. We don't know the
exact expression for that, however we can obtain an upper bound for 
its mass squared.
First, from the potential $V_{\rm tachyon}$, one can see that its mass
squared should be greater than $-1$. Second, since we know that it is
the lowest mode of the Schr\"odinger equation (\ref{flucL}), we can
use the following inequality for any smooth function $f$ 
(with appropriate boundary conditions)
\begin{eqnarray}
  m^2 \leq \frac{
\displaystyle\int \! dx \; f \left(
-\p_x^2 + V_{\rm tachyon}(\T)
\right) f
 }{
\displaystyle\int \! dx \; f^2 }.
\end{eqnarray}
{}From the observation above, we know that the true fluctuation is
almost a constant function everywhere, thus a constant function is
actually a good test function. Substituting this with noting that 
one has to perform the field redefinition (\ref{redef}), one obtains
\begin{eqnarray}
  m^2 \leq \frac{
\displaystyle\int \! dx \; {\rm e}^{-\T^2/2} 
\left(
-\p_x^2 + V_{\rm tachyon}(\T)
\right) {\rm e}^{-\T^2/2} 
 }{
\displaystyle\int \! dx \; {\rm e}^{-\T^2}  }.
\end{eqnarray}
Using the equations of motion, the numerator of the 
right hand side reads
\begin{eqnarray}
\displaystyle\int \! dx \; {\rm e}^{-\T^2} 
\left( (\p_x \T)^2 -1 
\right)
=
-2x_1 {\rm e}^{-T_0^2},
\end{eqnarray}
where at the last equality we have used the relation
(\ref{relation}). Since this is negative, it is shown that the system
admits a tachyonic fluctuation, as expected. This result can be derived 
without referring
to the special form of $\K'$ (\ref{bsftpot}). Only if the parameter
$A$ in (\ref{relation}) is non-zero (not the BPS limit), the above
estimation gives a tachyonic fluctuation. 

The constant test function is a better approximation for a larger
separation of the branes.  Therefore, the mass squared of this tachyonic
fluctuation can be estimated in the large brane separation as
\begin{eqnarray}
  m^2\sim -\frac{2}{\sqrt{\pi}}|T_0| {\rm e}^{-T_0^2}.
\label{evalu}
\end{eqnarray}
For a very large separation of the branes,
we can see that this state is nearly massless. If we take the BPS
limit ($|T_0| \rightarrow \infty$), the above state becomes exactly
massless, which is consistent with the fact that this limit is actually
separating the brane and the antibrane to the opposite spatial
infinities and turning off the inter-brane interactions to reproduce
two sets of the spectra of a single BPS brane. 
We will show later in Sec.\ \ref{sc:susybreaking} 
that the mass squared of tachyon (\ref{evalu}) can also 
be obtained by means of the low-energy theorem associated 
with the supersymmetry breaking. 

In terms of string theory, 
the possible interpretation of the tachyonic state with the mass
squared (\ref{evalu}) is given as follows. The matter content of the 
coincident brane-antibrane system consists of two transverse 
scalar fields $X^{(1)}$ and $X^{(2)}$, 
two gauge fields and a comlex tachyon field $\varphi$. Here we
assumed for simplicity a single transverse direction. The tachyon
field $\varphi$ becomes massive if the brane separation is larger than the 
string scale. Note that this massive tachyon field is not directly related 
with the tachyon fields which we have discussed, $T$ and $t$. 
There is the following coupling between the scalar fields $X$ 
and the massive tachyon $\varphi$ \cite{Hashimoto:2002xt}, 
\begin{eqnarray}
  (X^{(1)}-X^{(2)})^2 |\varphi|^2 \exp(-|\varphi|^2).
\end{eqnarray}
Since we are looking at the low energy excitation, the massive
$\varphi$ may be integrated out, then the above coupling gives a mass
term for the linear combination $X^{(1)}-X^{(2)}$. For the large
separation, the mass of $\varphi$ becomes large, and the integration
gives only small contribution, thus the mass generated from this term
is expected to be very small. This is expected to give the expression
(\ref{evalu}), since in the above coupling there appears the
exponential, and the typical mass scale of $\varphi$ is $\sim x_1 \sim 
T_0$. 
On the other hand, another linear combination
$X^{(1)}+X^{(2)}$ has no coupling like above, therefore the mass is
not generated by the integration and remains massless. This 
remaining massless mode is identified with the NG mode in our
fluctuation analysis, the antisymmetric combination of the flat
configurations for $t$.

\subsection{Fermion fluctuation}

It is known that on the non-BPS branes there are fermions which 
couple to the tachyon. The leading order fermion terms are given by
\cite{MZg}
\begin{eqnarray}
  S = -\int\! d^{10}x\; \left(\K'(T)\right)^2
\left[
\frac{i}2 \bar{\psi} \Gamma^\mu \p_\mu \psi +W(T)
\bar{\psi}\psi
\right],
\label{fermi}
\end{eqnarray}
where the Yukawa coupling is supplied with
\begin{eqnarray}
W(T) \equiv -\frac{\K''(T)}{\K'(T)} .
\end{eqnarray}
In this subsection we adopt $p=8$ for definiteness.
For the fermion spectrum in the tachyon solution background, we follow
the analysis \cite{MZg}. Performing the field redefinition
\begin{eqnarray}
  \chi \equiv \K'(\T) \psi
\end{eqnarray}
and decomposing the fields and the gamma matrices as
\begin{eqnarray}
\chi = 
\left(
    \begin{array}{cc}
\chi_1 \\
\chi_2
    \end{array}
\right) ,
\quad
  \Gamma^x = \left(
    \begin{array}{cc}
0 & iI \\
iI & 0
    \end{array}
\right),
\quad
  \Gamma^\mu = 
\left(
    \begin{array}{cc}
\gamma^\mu & 0 \\
0 & \gamma^\mu
    \end{array}
\right) \quad (\mu = 0,\cdots,8) ,
\end{eqnarray}
we have the Schr\"odinger equation for $\chi_\pm \equiv \chi_1 \pm
\chi_2$ as
\begin{eqnarray}
  \left(
-\p_x^2 + V_{\rm fermion}^{(\pm)}(\T)
\right)\chi_\pm
= m^2_\pm \chi_\pm.
\label{fere}
\end{eqnarray}
where the potential is given as 
\begin{eqnarray}
V^{(\pm)}_{\rm fermion} 
= (W(\T))^2 \pm \p_x W(\T)
= \left(\frac{\K''}{\K'}\right)^2
\mp \left[\frac{\K'''}{\K'}-
\left(\frac{\K''}{\K'}\right)^2\right]\p_x \T.
\end{eqnarray}
For the BSFT potential (\ref{bsftpot}), 
$V^{(\pm)}_{\rm fermion} = \T^2 \pm \p_x \T$.
It immediately follows that the mass squared $m_\pm^2$ is positive
semidefinite, since 
\begin{eqnarray}
 m_\pm^2 
= \frac{
\displaystyle\int\! dx \; 
\bar{\chi}_\pm \! \left(
-\p_x^2 + W(\T)^2 \pm \p_x W(\T)
\right)\!\chi_\pm
}{
\displaystyle\int\! dx \; \bar{\chi}_\pm \chi_\pm}
= \frac{
\displaystyle\int\! dx \; 
\left[\left(
\p_x \mp W
\right)
\!\bar{\chi}_\pm  
\right]
\left(
\p_x \mp W
\right)
\!\chi_\pm
}{
\displaystyle\int\! dx \; \bar{\chi}_\pm \chi_\pm}
\geq 0.
\nonumber
\end{eqnarray}
Moreover, from this inequality 
we can show that there exist massless modes. The massless modes 
satisfy the first order equations
\begin{eqnarray}
\left(
\p_x \mp W(\T)
\right)
\chi_\pm =0.
\end{eqnarray}
This is easily solved as
\begin{eqnarray}
 \chi_\pm \; \propto\; \exp\left[
\pm \int^x\!dx\; W(\T) 
\right].
\label{zerof}
\end{eqnarray}
For the BSFT potential (\ref{bsftpot}), we have $W(\T)=\T$, thus 
one can readily see that $\chi_+$ is localized on the brane 
located at $x=x_1/2$, while the other massless mode for $\chi_-$
is on the antibrane at $x=-x_1/2$. The localized wave functions are
almost Gaussian, $\chi_\pm \sim \exp (-(x\mp x_1/2)^2)$.

One can show that 
the mass spectra of $\chi_+$ and $\chi_-$ are identical, 
except for the massless modes. 
 Suppose that we have a 
solution for $\chi_+$ in the eigen equation (\ref{fere}). Then
constructing a new function
\begin{eqnarray}
 \widetilde{\chi}_- \equiv (-\p_x + W(\T)) \chi_+, 
\end{eqnarray}
it is easy to show that this $\widetilde{\chi}_-$ satisfies the $\chi_-$
eigen equation (\ref{fere}) with the mass $m_+^2$. This implies that for
each eigen mode $\chi_+$, there is an eigen state $\chi_-$ with the same
mass, provided $\chi_+$ is not a zero mode (then $\tilde \chi =0$). 
The converse can be shown in the same manner, therefore we have 
degeneracy of the fermion spectra between $\chi_+$ and $\chi_-$ 
except for the massless modes. 

Since in our case the world volume is compactified, the wave functions
are all normalizable. 
Hence both of the zero modes (\ref{zerof}) are consistent
solutions. Thus we conclude that 
the fermion spectrum 
contains one pair of zero modes and (infinitely) many pairs 
of massive modes. 
Namely fermions are doubly degenerate 
on our non-BPS brane-antibrane background 
even including massless modes. 

Another way to see this degeneracy is as follows. For non-BPS
configurations ($T_0 \neq \infty$) the solution we obtained has a period 
$2x_1$, and satisfies
\begin{eqnarray}
 \T(x+x_1) = -\T(x).
\end{eqnarray}
Hence 
this translation along $x$ gives an exchange 
$V^{(+)} \leftrightarrow V^{(-)}$. This provides an 
interesting relation between the eigen functions
\begin{eqnarray}
 \chi_\mp (x+x_1) = \chi_\pm(x),
\end{eqnarray}
Therefore it is a direct result of this relation that the spectrum is
doubly degenerate. This argument is valid even for the massless modes.
However, for the BPS case (a single D-brane
solution), there is no periodicity and thus the degeneracy cannot be
observed in this way. In fact, the degeneracy is only for the massive
modes in this BPS case \cite{MZg}. One of the zero modes (\ref{zerof})
becomes non-normalizable.

\subsection{Gauge fluctuation}

It is possible to analyse the gauge fluctuation in the same manner. 
The Lagrangian we consider is
\begin{eqnarray}
{\cal L} =  -\left(\K'(T)\right)^2
 \left(1 + \partial_\mu T \partial^\mu T 
 + \frac14 F_{\mu\nu}F^{\mu\nu}\right).  
\end{eqnarray}
For the two-derivative truncation of the BSFT, we have (\ref{bsftpot})
\cite{MZg,Terashima:2001va}.
Performing the same analysis given in \cite{MZ,MZg} leads to the
following Schr\"odinger equation
\begin{eqnarray}
 \left(
-\p_x^2 + V_{\rm gauge}(\T)
\right) \hat{A}_{\hat{\mu}} = m_{\rm gauge}^2 \hat{A}_{\hat{\mu}}.
\end{eqnarray}
where $\hat{A}_{\hat{\mu}} \equiv {\cal K}'(T_{\rm cl}){A}_{\hat{\mu}}$ 
and the potential is given by 
\begin{eqnarray}
 V_{\rm gauge}(\T)\equiv \left(\frac{\K''}{\K'}\right)^2
+\left[  \frac{\K'''}{\K'}-\left(\frac{\K''}{\K'}\right)^2
\right](\p_x \T )^2.
\end{eqnarray}
The important point is that the operator in the Schr\"odinger equation can be
recast into a form
\begin{eqnarray}
 -\p_x^2 + V_{\rm gauge}(\T) = D^\dagger D
\label{rewre}
\end{eqnarray}
where the linear differential operator $D$ and $D^\dagger$ 
are defined as
\begin{eqnarray}
 D \equiv i \left(\p_x - \frac{\K''}{\K'} (\p_x \T)\right)
= i \K' \p_x\frac{1}{\K'}, 
\quad 
 D^\dagger \equiv -i \left(\p_x + \frac{\K''}{\K'} (\p_x \T)\right)
= i \frac{1}{\K'} \p_x\K'. 
\label{D}
\end{eqnarray}
These operators $D$ and $D^\dagger$ are adjoint of each other 
with respect to the inner product for $\hat{A}_{\hat \mu}$ fields 
$(\phi_1(x), \phi_2(x)) \equiv \int dx \phi_1(x) \phi_2(x)$, as 
\begin{eqnarray}
(\phi_1(x), D\phi_2(x)) = (D^\dagger \phi_1(x), \phi_2(x)).
\end{eqnarray}
Defining the eigen function $u_n(x)$ as $D^\dagger D u_n = m_n^2 u_n$, 
we obtain from the expression (\ref{rewre}) 
that the mass 
squared $m^2_{\rm gauge}$ is positive semi-definite 
\begin{eqnarray}
 m^2_{n} (u_n, u_n) = 
 (u_n, D^\dagger D u_n) = (D u_n, D u_n) \geq 0.
\end{eqnarray}
This implies that the massless mode should satisfy a linear
differential equation 
\begin{eqnarray}
 D u_0(x) = 0  \qquad {\rm for} \quad m_0 =0.
\end{eqnarray}
From the last expression in (\ref{D}), this massless mode can be easily
found as 
\begin{eqnarray}
 u_0(x) = \K'(\T(x)).
\end{eqnarray}
Therefore we have found that the mass spectrum of the gauge fluctuations
consists of one massless mode and massive modes, thus there is no
tachyonic state.

\subsection{Supersymmetry breaking}
\label{sc:susybreaking}

Although there are massless modes for all sectors  of the fluctuation 
(tachyon (scalar), fermion and gauge fields), 
a tachyonic mode exists only
in the tachyon (scalar) fluctuation $t$, as shown in (\ref{evalu}). 
This is a reflection of the fact that the supersymmetry is broken for our 
solution. The amount of the supersymmetry breaking is measured by the 
difference of the boson/fermion spectra, that is (\ref{evalu}) for the 
BSFT potential (\ref{bsftpot}). 

The massless mode of the tachyon fluctuation is the NG boson associated
with the breaking of the translation invariance. Analogously, 
the massless modes of the fermion fluctuation can be thought of
as NG fermions coming from the breaking of the overall target space 
supersymmetries as follows. 

Non-BPS branes in superstring theory are unstable (meta-stable) 
and the associated
tachyon $T$ develops a vacuum expectation value. 
To describe this tachyon condensation, an effective field 
theory of tachyons away from $T=0$ has been proposed. 
Since this field theory is originally a field theory on the non-BPS 
branes, the supersymmetries are completely broken at $T=0$ 
and are realized only nonlinearly \cite{Terashima:2001va}. 
This is the target space supersymmetries of $1+9$ dimensions. 
In the limit of tachyon condensation $T\rightarrow \pm \infty$, 
supersymmetry is recovered. 
If we have a linear solution (\ref{linear}) 
of the effective field theory corresponding to a single BPS brane, 
half of the supersymmetries is recovered and is linearly realized. 
On the other hand, the brane-antibrane configuration breaks this 
recovered linearly supersymmetries.
Therefore we should have the NG fermion 
associated with the broken target space supersymmetries. 

Let us elaborate the above consideration more precisely. 
The total configuration of the brane and antibrane 
breaks supersymmetries completely. 
The amount of supersymmetry breaking is 
measured by the mass squared of the tachyon fluctuation
(\ref{evalu}). There is another way to observe this. The non-linearly
realized supersymmetries of the original action 
with the fermions (the sum of (\ref{Lag}) and (\ref{fermi}))
are given as \cite{Terashima:2001va}
\begin{eqnarray}
 \delta \psi = \epsilon - (\p_\mu T) \Gamma^\mu \epsilon.
\end{eqnarray}
Therefore one can see that, for the linear solution (\ref{linear}), 
half of the nonlinear supersymmetries are restored to be linear
supersymmetries. This shows that the kink is a BPS D-brane. 
In our case, near the branes $x=\pm x_1/2$ the solution is approximated
by a linear function $T\sim \mp q (x\pm x_1/2)$ (\ref{approx}), where
$q\sim 1 + {\cal O}({\rm e}^{-T_0^2})$. 
Plugging this into the above supersymmetry
transformation, we find that it is slightly broken by 
${\rm e}^{-T_0^2}$.  
This is consistent with the fact that in the fluctuation spectra 
the tachyon mass squared is of order ${\cal O}({\rm e}^{-T_0^2})$, 
as shown in (\ref{evalu}). 

\begin{figure}[tdp]
\begin{center}
\begin{minipage}{70mm}
\begin{center}
   \leavevmode
   \epsfxsize=70mm
   \epsfbox{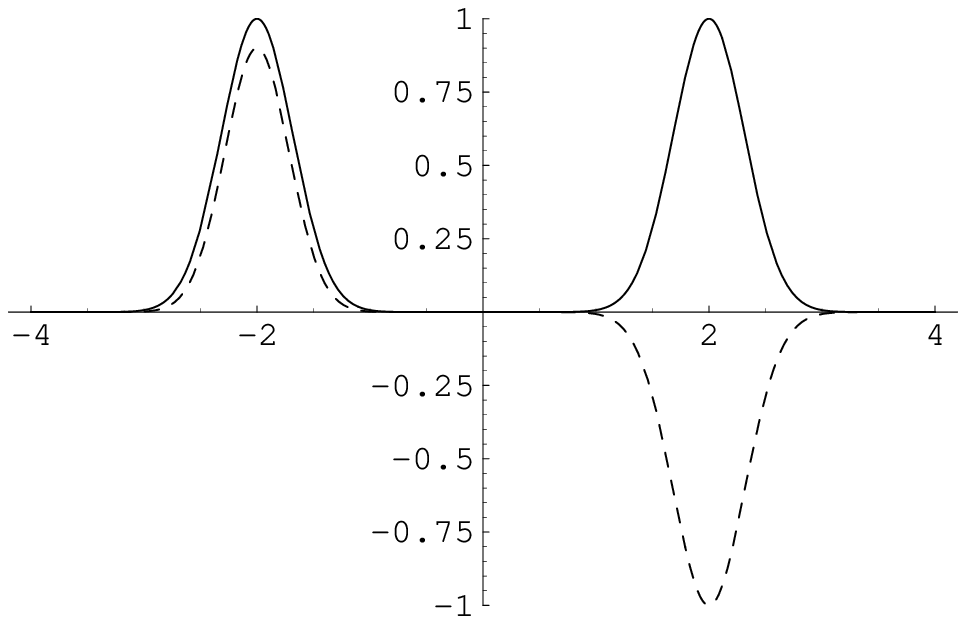}
\put(0,60){$x$}
\put(-60,40){$\displaystyle\frac{x_1}{2}$}
\put(-100,130){$\hat{t}$}
\put(-95,65){$0$}
\caption{Two lowest fluctuations of the tachyon field $T$. 
The solid line denotes the tachyonic fluctuation, while the dashed line
is the massless NG mode. 
}
\label{boson}
\end{center}
\end{minipage}
\hspace{7mm}
\begin{minipage}{70mm}
\begin{center}
   \leavevmode
   \epsfxsize=70mm
   \epsfbox{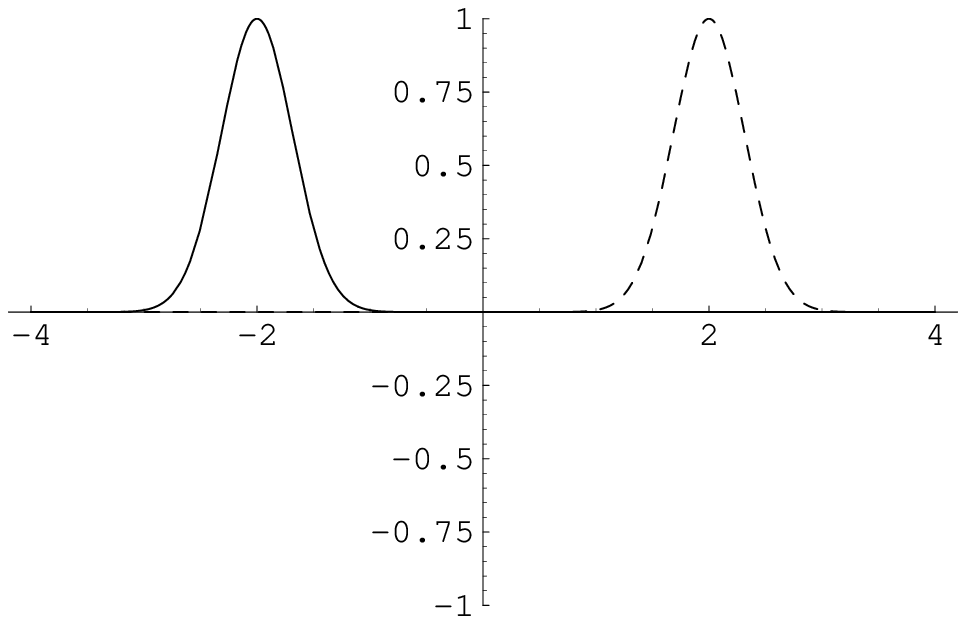}
\put(0,60){$x$}
\put(-60,40){$\displaystyle\frac{x_1}{2}$}
\put(-160,40){$-\displaystyle\frac{x_1}{2}$}
\put(-60,130){$\chi_+$}
\put(-150,130){$\chi_-$}
\put(-95,65){$0$}
\caption{Fermionic fluctuation $\chi_{\mp}$ localized at 
$\mp x_1/2$ is shown
 by a solid (dashed) line. 
}
\label{fermion}
\end{center}
\end{minipage}
\end{center}
\end{figure}

Another consistency check can be made through 
the low-energy theorem associated with the supersymmetry 
breaking \cite{Maru:2000sx,Maru:2001sx}. 
Since the linearly realized supersymmetry of the tachyon condensed 
vacuum ($T=\pm \infty$) is spontaneously broken by the brane-antibrane 
configuration, low-energy theorems apply for the coupling of 
the NG fermion. 
When supersymmetry is spontaneously broken, the supercurrent $J^\mu$ 
at vanishing momentum reduces to the NG fermion field $\psi_{\rm NG}$ 
\begin{equation}
 J^{\mu}_{\alpha}
=\sqrt{2}if\cdot\left(\gamma^{\mu}\psi_{\rm NG}\right)_{\alpha}
 +J^{\mu}_{{\rm matter},\alpha}+\cdots, 
 \label{super_cc_3D}
\end{equation}
where $f$ is the order parameter of the supersymmetry breaking and 
$J^{\mu}_{{\rm matter},\alpha}(x)$ is the supercurrent for matter
fields. The conservation of the supercurrent $\partial_\mu J^\mu_{\alpha}=0$
relates the matrix elements of NG fermion to that of the matter
supercurrent. If we evaluate them between boson and fermion states of a
supermultiplet, the matrix element of the NG fermion gives the effective
three point coupling 
$g_{\rm eff}$ of the NG fermion with the supermultiplet of 
boson and fermion, whereas that of matter supercurrent gives 
the difference of squared mass 
$\Delta m^2=m_{\rm boson}^2-m_{\rm fermion}^2$ 
between the boson and the fermion. 
The resulting low-energy theorem is \cite{Maru:2001sx}
\begin{equation}
 g_{\rm eff}{}=\frac{\Delta m^2}{f}. \label{GTR-hef}
\end{equation}
The order parameter of the supersymmetry breaking $f$ is given by 
the energy $V_0$ of the background solution as 
$f=\sqrt{V_0}$. 
The effective three point coupling can be given by the 
overlap integral of the three wave functions in $x$. 
If we take the massless fermion localized on the brane as the matter 
fermion, 
its superpartner is given by the boson localized on our brane 
which is approximately given by an equal weight superposition of the 
zero mode and the tachyonic mode as shown in Fig.\ \ref{boson}. 
On the other hand, the NG fermion is localized on the 
antibrane as shown in Fig.\ \ref{fermion}. 
The three point coupling is given in terms of an overlap integral 
of the wave functions of the supermultiplet of the boson and 
fermion with the wave function of the NG fermion 
which can be efficiently evaluated by the single 
kink approximation \cite{Maru:2000sx}. 
Since the wave function of the NG fermion is localized 
at the antibrane and has an exponentially 
suppressed tail around the brane, 
the overlap integral is exponentially suppressed to give the order
${\cal O}({\rm e}^{-T_0^2})$ as a function of $T_0$ 
in agreement with our result of the direct evaluation of 
the tachyon mass squared in (\ref{evalu}).

It should be noted that the potentials for each fluctuation, 
$V_{\rm tachyon}$, $V_{\rm fermion}^{(\mp)}$, and $V_{\rm gauge}$,
coincide with each other if the classical solution is that of the
single brane, $\T=\pm (x-x_0)$. Therefore in the BPS limit our result
reproduces that of \cite{MZg}, however we note
that in our computation the derivation of the potential does not 
refer to any particular expression of the solutions. 

In sum, if the solution deviates from the linear case ($\T=x$), 
the 
potentials for each fluctuation differ, and generically the degeneracy
of the boson/fermion/gauge fluctuation spectra is resolved. For any
classical background solution one finds as fluctuations 
a massless bosonic scalar mode (NG boson), 
two fermionic modes which are NG fermions, and a massless vector 
mode.\footnote{This massless vector mode is supposed to be confined after
the annihilation of the brane and the antibrane (see for example,
\cite{yi, sen3}).}


\section{Brane-antibrane annihilation}

\begin{figure}[htbp]
\begin{center}
\begin{minipage}{100mm}
\begin{center}
   \leavevmode
   \epsfxsize=100mm
   \epsfbox{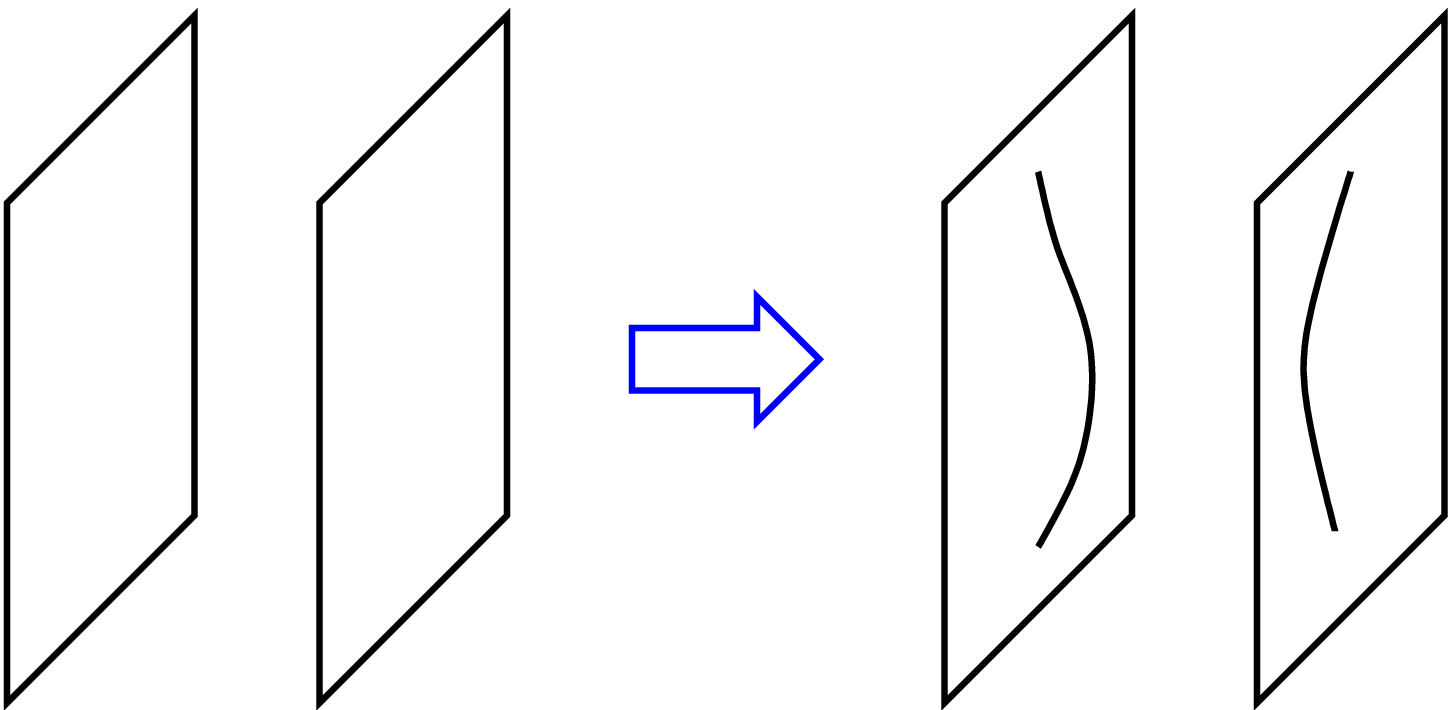}
\put(-200,0){$T<0$}
\put(-260,0){$T>0$}
\put(-330,0){$T<0$}
\caption{The surfaces of the brane and antibrane are deformed, and they
 start approaching. 
}
\label{fig2}
\end{center}
\end{minipage}
\end{center}
\end{figure}

Since there is a tachyonic fluctuation, the obtained solution 
is meta-stable and decays by some quantum effect. 
In this section we describe how the constructed solution of the 
brane and antibrane decays. 
The most natural decay may be 
by approach of the brane and antibrane.
As we noted, the vacuum expectation value (vev) 
of the tachyonic fluctuation
corresponds to the distance between the brane and antibrane. 
Thus this decay  mode is actually the rolling down of the potential hill
of the tachyonic fluctuation mode. 
From the mass squared of it (\ref{evalu}), one can write the potential
around the vacuum as 
\begin{eqnarray}
 V(t) \sim \frac{-T_0 {\rm e}^{-T_0^2}}{2}\; t^2 + {\cal O}(t^4).
\end{eqnarray}
The rolling down of the potential hill starts from quantum fluctuation, 
and a quantum-field-theoretical treatment for this kind of decay 
has been known for years
\cite{Guth:1985ya,Weinberg:1987vp}.\footnote{Application for the string
theory context, see \cite{Marcus:1988vs}.
Discussion in the modern D-brane context, see
\cite{Bardakci:2001ck,Craps:2001jp,Andreev:2001ak}.
}
Quantum one-loop corrections to the potential gives an imaginary part,
and that is regarded as a decay width of the system. In our case, the
world volume theory is $p+1$ dimensional, then the decay width for a
unit worldvolume can be obtained by following the computation in
\cite{Weinberg:1987vp,Craps:2001jp} as
\begin{eqnarray}
 \Gamma = \frac{\pi}{\left(\frac{p+1}{2}\right)!} 
\left(
\frac{-m^2}{4\pi}
\right)^{\frac{p+1}{2}}.
\end{eqnarray}
Here we derived this for the case of odd $p$ for simplicity. 
$m^2$ is the tachyon mass which is now given by (\ref{evalu}), therefore
the rough estimate of the decay width is 
\begin{eqnarray}
 \Gamma \sim \exp \left( - \frac{p+1}{2}T_0^2\right).
\end{eqnarray}
This decay width describes the approaching of the branes to each other,
as shown in Fig.\ \ref{fig2}.
The size of the supporting domain of the deformed surface can be also
estimated \cite{Weinberg:1987vp} as 
\begin{eqnarray}
 r_0 \sim 1/m \sim \exp (T_0^2).
\end{eqnarray}

For the brane and antibrane, there is another interesting decay
mode. This is 
nucleation of a throat connecting the two branes 
\cite{Callan:1997kz,Hashimoto:2002xt}.
The decay starts with a quantum tunnel effect by the nucleation of the
throat, and then the throat expands to sweep out the brane and antibrane.  
The decay width can be easily computed by use of the tension of the
branes and the brane separation.  
The radius of the nucleated throat is given as \cite{Hashimoto:2002xt} 
\begin{eqnarray}
 r = \frac{p}{2} x_1
\end{eqnarray}
where we used that $x_1\sim 2T_0$ is the brane separation. 
Since the decay width is given by the action
of the throat (which equals the tension times the area of the throat), 
\begin{eqnarray}
 \Gamma \sim \exp \left( -T_0^{p+1}\right). 
\end{eqnarray}
Therefore this is not a dominant decay mode if $p>1$. When $p=1$, this
decay process is comparable with the first one (gradual approaching).

For both decay modes, after the annihilation of the brane and antibrane,
we will be left 
with an almost homogeneous tachyon vev $T\sim T_0$. This is
unstable and rolls down the original potential ${\rm e}^{-T^2}$, and
finally we will be led to the rolling tachyon phase \cite{Senroll}.

\begin{figure}[tdp]
\begin{center}
\begin{minipage}{100mm}
\begin{center}
   \leavevmode
   \epsfxsize=100mm
   \epsfbox{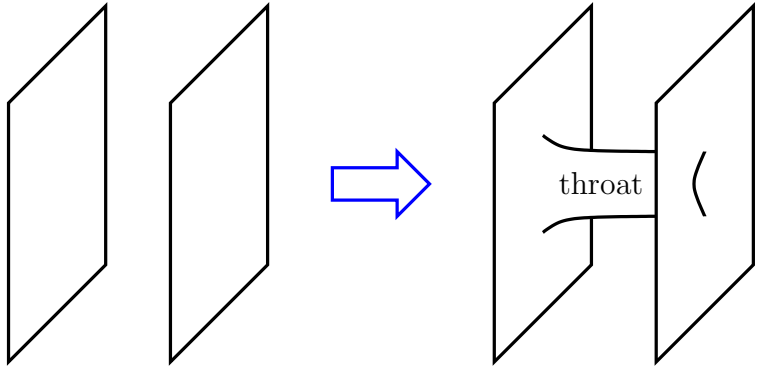}
\put(-75,65){throat}
\caption{Nucleation of a throat connecting the brane and the antibrane. 
}
\label{fig3}
\end{center}
\end{minipage}
\end{center}
\end{figure}

\section{Conclusion and discussion}

In this note we have constructed a classical solution representing a
brane-antibrane pair in a compact spacetime, in the tachyon effective
field theory known as Minahan-Zwiebach model. 
The force acting between the
branes comes from the modes propagating between the brane
and the antibrane. The mass of this mode 
is shown to agree with the 
excitation of the string stretched between the branes. Fluctuation
spectra for the tachyon field, the fermions and the gauge fields are
obtained, and they are found to exhibit the supersymmetry breaking
structure. There are two  possible decay modes of this configuration,
one is the approach of the branes to each other, and the other is the
nucleation of a throat connecting the branes. It is shown that the
former decay mode is dominant in our theory.

Although the tachyon effective theory considered here has many desirable
properties, it may receive $\alpha'$ corrections in string theory, and
there is no guarantee that the solution considered here survives. In
fact, even for the linear tachyon profile of a single D-brane solution,
the corresponding 
BSFT action includes infinitely many higher terms. A discussion on
this higher derivative treatment is briefly given in Appendix
A. However, we point out here that our solution is equipped with a
property which coincides with the solution of full string theory -- a
conformal field theory construction of brane and antibrane given by
A.\ Sen \cite{sen2}. There exists an exactly marginal deformation of the CFT
which is a form of $\lambda \sin(X)$ where $X$ is a scalar field in two 
dimensional worldsheet theory \cite{cos}, and $\lambda$ is a deformation
parameter. This deformation is linear around  
$X\sim 0$ while $\sim X^2$ around the turning point (in our solution,
around $T_0$). Since this deformation is exactly marginal, the
configuration is a solution of full string theory at the tree level. 
Therefore an optimistic view is that our solution is not far from the
full solution. It is important to construct an off-shell formalism
around this marginally deformed background.\footnote{See \cite{Sachs}
  for the related study in the context of tachyon effective theories.
In Berkovits' formulation of
superstring field theory, similar classical solutions
were constructed in \cite{Ohmori}.}

The brane-antibrane interaction in string theory should be dictated by
the exchange of gravitons and RR gauge fields. However in our setup we
neglected these closed string modes from the first place, thus we cannot
see this in the force estimation in section 3. In one sense this was
good for us since we wanted to see the open string off-diagonal mode. 
However for the precise description of the brane and antibrane the lack of
gravitational interaction is fatal, and how the closed string effect is
incorporated in the tachyon  model should be studied, for the precise
description of the decay for example.

Our brane-antibrane solution can be formed dynamically, if we start from
an unstable homogeneous vacuum $T=0$. 
If the tachyon starts rolling inhomogeneously,
generically the direction of the rolling may differ in regions of the
world volume, and a stack of brane-antibrane pairs may be formed. The
time-dependent inhomogeneous decay of tachyon theory was 
recently studied by several people \cite{several,Hashimoto:2002xt} in
various approaches. Here again the inclusion of the closed string modes
may be concern. We leave these issues to the future work.

\vspace{10mm}

\noindent
{\Large \bf Acknowledgment}

The authors thank the Yukawa Institute for Theoretical Physics at Kyoto
University, where a part of this work was done during the conference 
``Quantum Field Theory 2002'' YITP-W-02-04. 
K.\ H.\ is grateful to S.\ -H.\ Henry Tye for useful discussions.
N.\ S.\  is supported in part by Grant-in-Aid for Scientific 
Research from the Ministry of Education, Science and Culture 
 13640269. 

\appendix

\section{BSFT treatment with higher derivative terms}

In the whole analysis in this paper, so far we have used the
Minahan-Zwiebach model (\ref{Lag}). Although this model is field
theoretically very interesting and shares common properties with 
string theory low energy behavior, from a strict point of view, it is
just a derivative truncation of the BSFT action \cite{Witten1, moore}
and receives higher
derivative corrections. In this section we demonstrate to what extent
the examined properties of the brane-antibrane solution survives in the
derivative corrections. 

\begin{figure}[htdp]
\begin{center}
\begin{minipage}{100mm}
\begin{center}
   \leavevmode
   \epsfxsize=80mm
   \epsfbox{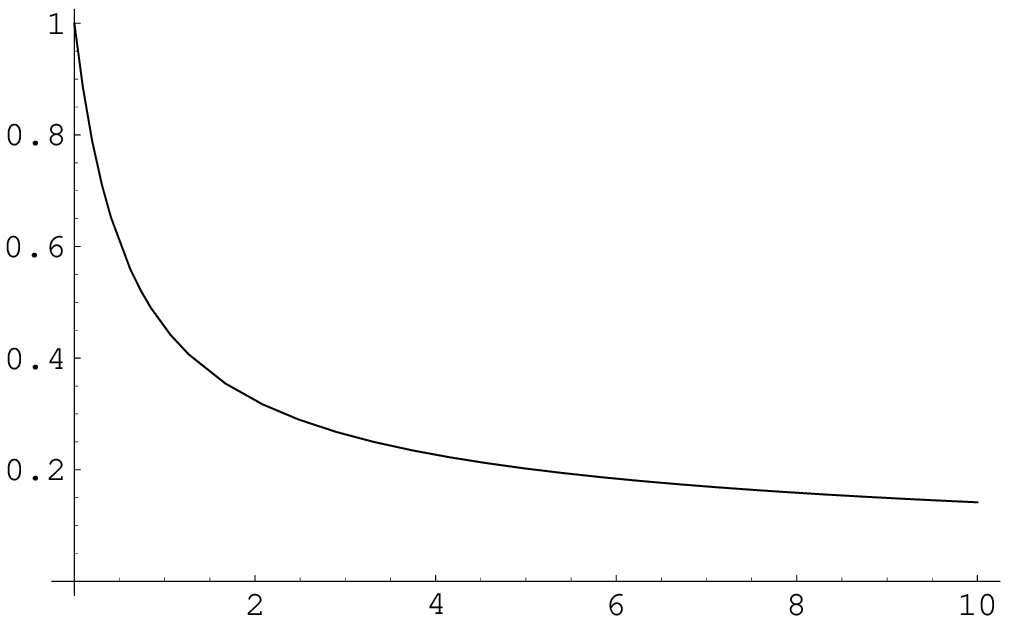}
\put(10,0){$y=(\p T)^2$}
\put(-210,150){${\cal F}(y)-2 y {\cal F}'(y)$}
\caption{The right hand side of the equation (\ref{bsre}). 
}
\label{bs}
\end{center}
\end{minipage}
\end{center}
\end{figure}

The BSFT action for the linear tachyon profile (equivalently, neglecting
the $\p \p T$ terms) is given in \cite{moore} as
\begin{eqnarray}
 S = -\int \! d^{p+2}x {\rm e}^{-T^2/4} {\cal F}(y),
\label{bsftac}
\end{eqnarray}
where 
\begin{eqnarray}
 {\cal F}(y) = \frac{y 4^y (\Gamma(y))^2}{2 \Gamma(2y)},
\quad
y \equiv (\p_\mu T )^2.
\end{eqnarray}
We can solve the equations of motion in a similar way, by constructing a
conserved quantity for the translational invariance, 
\begin{eqnarray}
 A = \frac{\delta L}{\delta \p T} \p T - L
= {\rm e}^{-T^2/4} \left({\cal F}-2 y {\cal F}'\right).
\end{eqnarray}
Therefore, the redefinition of the constant $A = {\rm e}^{-T_0^2/4}$ leads to
\begin{eqnarray}
 {\rm e}^{(T^2 - T_0^2)/4} = {\cal F}-2 y {\cal F}'.
\label{bsre}
\end{eqnarray}
The MZ model has ${\cal F}(y)=1+y$. Therefore the right hand side of the
above equation is linear, ${\cal F}-2 y {\cal F}'=1-y.$ 
It becomes zero at $y=1$, and this indicates that the solution 
at $T=0$ for large $|T_0|$ becomes almost linear with slope $|\p T| =1$. 
On the other hand, in the BSFT case, the right hand side has no zero
(see Fig.\ \ref{bs}). The limit ${\cal F}-2y{\cal F}' \rightarrow 0$
corresponds to $y \rightarrow \infty$. 
This means that, the slope $q$ at which $T=0$ (the location of
the branes) is very large. This slope $q$ diverges if we take the BPS
limit $T_0=\infty$, which is consistent with the previous analyses in
the BSFT.

As before, the estimation of the slope $q$ may lead to the evaluation of
the brane-antibrane separation. In the asymptotic region of $y$ we can
approximate the right hand side of (\ref{bsre}) as
\begin{eqnarray}
 {\cal F}-2 y {\cal F} \sim 
\sqrt{\frac{\pi}{16 y}} + {\cal O}\left(\frac{1}{y^{3/2}}\right).
\end{eqnarray}
Therefore for large $|T_0|$, the slope at $T=0$ can be approximated by
the solution
\begin{eqnarray}
 -\frac14 T_0^2 \sim -\frac12 \log y.
\end{eqnarray}
Since in this linear approximation the inter-brane separation $x_1$ is
given by $x_1= T_0/q$, we substitute $y(T=0) = q^2 = (T_0/x_1)^2$ into
the above equation and obtain
\begin{eqnarray}
x_1 \sim |T_0| {\rm e}^{-T_0^2/4}. 
\end{eqnarray}
Unfortunately, this means that the large $|T_0|$ does not corresponds to
the large separation. Hence our linear approximation is invalid. 
However, it is obvious from Fig.\ \ref{bs} that a
similar oscillatory solution exists and exhibits periodicity.
Since we have not found a suitable limit of the parameters for the large
brane separation, it seems to be difficult to see the structure of the
supersymmetry breaking without performing explicit fluctuation analysis
using the solutions, although it is obvious that the supersymmetry
breaking occurs. We expect similar processes for the brane-antibrane 
annihilation. 

Though here we have considered only the higher derivative corrections
of the form $(\partial T)^n$ $(n\geq 4)$, 
inclusion of the terms with higher
derivatives acting on T, such as $\partial^2 T$, 
would be also important.
The relevant action was analysed in \cite{Tsey}, and it would be
interesting by using the action in \cite{Tsey} 
to see if the brane-antibrane
solution survives also these corrections.\footnote{
We would like to thank A. Tseytlin for bringing us this viewpoint.}

\newcommand{\J}[4]{{\sl #1} {\bf #2} (#3) #4}
\newcommand{\andJ}[3]{{\bf #1} (#2) #3}
\newcommand{\AP}{Ann.\ Phys.\ (N.Y.)}
\newcommand{\MPL}{Mod.\ Phys.\ Lett.}
\newcommand{\NP}{Nucl.\ Phys.}
\newcommand{\PL}{Phys.\ Lett.}
\newcommand{\PR}{ Phys.\ Rev.}
\newcommand{\PRL}{Phys.\ Rev.\ Lett.}
\newcommand{\PTP}{Prog.\ Theor.\ Phys.}
\newcommand{\hep}[1]{{\tt hep-th/{#1}}}

\end{document}